\documentclass[iop]{emulateapj}
\usepackage{apjfonts}
\journalinfo{The Astrophysical Journal Letters, 2014, in press}

\shorttitle{Suprathermal Electrons in the Solar Corona}
\shortauthors{S.~R.~Cranmer}

\begin{document}

\title{Suprathermal Electrons in the Solar Corona:
Can Nonlocal Transport Explain Heliospheric Charge States?}

\author{Steven R. Cranmer}
\affil{Harvard-Smithsonian Center for Astrophysics,
60 Garden Street, Cambridge, MA 02138, USA}

\begin{abstract}
There have been several ideas proposed to explain how the Sun's
corona is heated and how the solar wind is accelerated.
Some models assume that open magnetic field lines are heated by
Alfv\'{e}n waves driven by photospheric motions and dissipated
after undergoing a turbulent cascade.
Other models posit that much of the solar wind's mass and energy
is injected via magnetic reconnection from closed coronal loops.
The latter idea is motivated by observations of reconnecting
jets and also by similarities of ion composition between closed
loops and the slow wind.
Wave/turbulence models have also succeeded in reproducing
observed trends in ion composition signatures versus wind speed.
However, the absolute values of the charge-state ratios predicted
by those models tended to be too low in comparison with observations.
This letter refines these predictions by taking better account of
weak Coulomb collisions for coronal electrons, whose thermodynamic
properties determine the ion charge states in the low corona.
A perturbative description of nonlocal electron transport is
applied to an existing set of wave/turbulence models.
The resulting electron velocity distributions in the low corona
exhibit mild suprathermal tails characterized by ``kappa''
exponents between 10 and 25.
These suprathermal electrons are found to be sufficiently energetic
to enhance the charge states of oxygen ions, while maintaining the
same relative trend with wind speed that was found when the
distribution was assumed to be Maxwellian.
The updated wave/turbulence models are in excellent agreement
with solar wind ion composition measurements.
\end{abstract}


\section{Introduction}
\label{sec:intro}

The existence of the solar wind, a continuous outflow of charged
particles from the Sun, is believed to be a direct result of the
heating of plasma to temperatures of order $10^6$~K in the
solar corona \citep{P58a}.
However, the physical processes responsible for the wind and the
corona have not yet been identified conclusively
\citep[see, e.g.,][]{Ma06,Cr09,PD12}.
Much of the heliospheric plasma is of sufficiently low density
to make particle--particle collisions infrequent.
This means that some aspects of particle distributions measured
in interplanetary space may carry information about the distant
coronal heating.
For example, the ionization states of most heavy ions are believed
to be ``frozen in'' low in the corona and remain constant between
heights of a few solar radii ($R_{\odot}$) and 1~AU.
Above a certain point in the solar atmosphere, the ions collide
with virtually no electrons and thus do not undergo any additional
ionization or recombination \citep{Hu68,Oe83}

Charge states measured at 1~AU have been used as indirect probes of
the near-Sun plasma, with coronal electron temperatures
$T_{e} \approx 1.5$~MK often being inferred \citep{Ge95,Ko97}.
However, spectroscopic measurements of temperature-sensitive
emission line ratios typically gave $T_{e} \lesssim 0.9$~MK
at the heights where freezing-in should take place.
\citet{EE00,EE01} found that this discrepancy may disappear if
either: (1) electrons have non-Maxwellian velocity distribution
functions (VDFs), or
(2) ions of different charge states flow with different speeds
in the corona.
More recent comparisons of spectroscopic and in~situ measurements
\citep[e.g.,][]{Ln12b,Ko14} continue to attempt to reconcile these
observations, but no single model has been found that explains
everything.

Although there is no direct evidence for ion--ion differential
streaming in the corona, there are hints that the electrons may
have non-Maxwellian VDFs.
Some earlier studies appeared to rule out the existence of
suprathermal electrons at low coronal heights \citep{An96,Ko96},
but the evidence may be starting to swing in the other
direction \citep[e.g.,][]{Ra07,Ku11}.
Theoretically, a suprathermal electron ``tail'' may be the
natural outcome of the dissipation of turbulent plasma
fluctuations \citep{RM98,Vi00,Yo06} or the ballistic
acceleration of coronal jets \citep{Fe12}.
\citet{Sc92} suggested that any small nonthermal tail in
the lower atmosphere can be amplified in the corona by
gravitational filtration \citep[see also][]{P58b,Lv74}.

Another possibly important source of non-Maxwellian VDFs may be the
nonlocal transport of electrons through regions of weak collisionality.
Because of the complex velocity dependence of Coulomb collisions,
the electron distribution at one heliocentric distance depends on
the properties of electrons over a range of surrounding distances.
\citet{OS78} suggested that suprathermal ``halo'' electrons
seen at 1~AU are likely to be the remnant of a hot coronal VDF.
\citet{SO79} provided a straightforward model---reminiscent of
radiative transfer in astrophysics---for estimating the magnitude
of these effects in the coupled corona--heliosphere system.

This letter explores the consequences of nonlocal electron transport
on the formation of the frozen-in charge states (e.g., O$^{+7}$
and O$^{+6}$) in an existing model of turbulent coronal heating
and solar wind acceleration.
\citet{CvB07} found that the assumption of Maxwellian electrons
resulted in values of the O$^{+7}$/O$^{+6}$ ionization state ratio
that were too low by about an order of magnitude in comparison to
observations.
Section \ref{sec:so79} applies the \citet{SO79} transport framework
to a representative fast-wind model.
Section \ref{sec:freeze} shows how mild suprathermal enhancements
at $r \approx 1.02 \, R_{\odot}$, that result from collisional
transport, appear to be sufficient to increase the frozen-in
ionization states to the observed levels.
Section \ref{sec:conc} summarizes the results and gives suggestions
for future improvements.

\section{Nonlocal Model of Electron Transport}
\label{sec:so79}

\begin{figure}
\epsscale{1.14}
\plotone{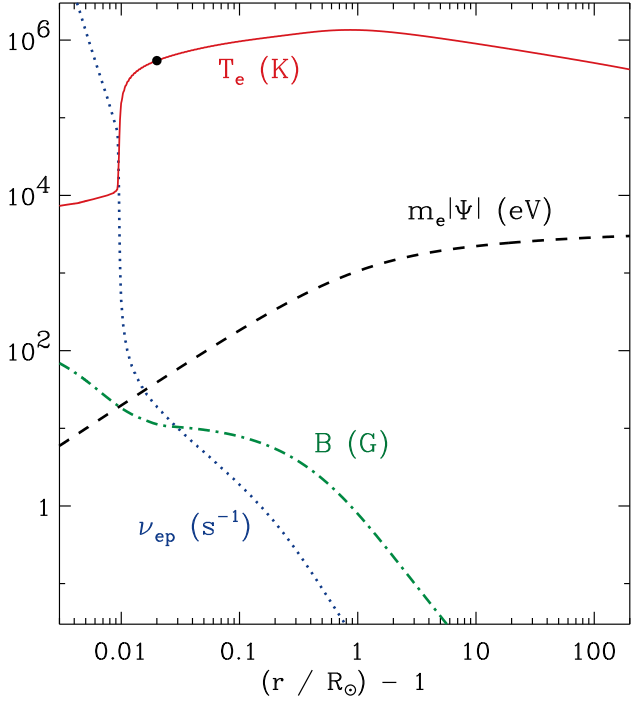}
\caption{Radial dependence of plasma parameters from the
\citet{CvB07} coronal hole model, including
$T_e$ (red solid curve),
$B$ (green dot-dashed curve),
$m_{e}|\Psi|$ (black dashed curve), and
$\nu_{ep}$ (blue dotted curve).
\label{fig01}}
\end{figure}

\citet{CvB07} presented steady-state solutions to the conservation
equations of mass, momentum, and energy for superradially expanding
flux tubes rooted in the solar photosphere.
The coronal heating in these models was produced self-consistently
via the inclusion of gradual Alfv\'{e}n-wave reflection and the
dissipation of magnetohydrodynamic (MHD) turbulence.
Here we use one-fluid plasma properties from the \citet{CvB07}
polar coronal hole model as proxies for the {\em{electron}}
density, flow speed, and temperature.
Figure \ref{fig01} shows the radial dependence of electron
temperature $T_e$ and magnetic field magnitude $B$ for this model.
The first-order assumption (to be perturbed below) is that the
electrons obey a locally Maxwellian VDF $f_{\rm M}(v)$ with a radially
dependent thermal speed $w_{e}=(2k_{\rm B}T_{e}/m_{e})^{1/2}$.

In order to use these electron properties as inputs to the
\citet{SO79} model, the radial variation of the charge-separation
electric field must be computed.
The electron momentum conservation equation allows us to estimate
the gradient of the electric potential $\Phi$
\citep{Jo70,Ho70}, with
\begin{equation}
  e\frac{\partial\Phi}{\partial r} \, = \,
  \frac{1}{n_{e}}\frac{\partial}{\partial r}
  \left( n_{e} k_{\rm B} T_{e \parallel}\right)
  +\frac{1}{B}\frac{\partial B}{\partial r} k_{\rm B}
  \left( T_{e \perp}-T_{e \parallel}\right) \,\, ,
  \label{eq:dPhidr}
\end{equation}
and this is simplified by assuming $T_{e\parallel}=T_{e\perp}=T_e$.
Equation (\ref{eq:dPhidr}) is integrated numerically to obtain
$\Phi(r)$.
The electric potential combines with gravity to give the total
potential felt by collisionless electrons,
\begin{equation}
  \Psi \, = \, -2 \left( \frac{GM_{\odot}}{r} +\
  \frac{e \Phi}{m_e} \right)
\end{equation}
and energy conservation is equivalent to the assumption that
the quantity $v_{\parallel}^{2} + v_{\perp}^{2} + \Psi$
remains constant along an electron's trajectory.
In combination with magnetic moment conservation
($v_{\perp}^{2}/B = \mbox{constant}$), this specifies the full
``history'' of an electron that ends up at a given radius $r$
with known velocity components $v_{\parallel}$ and $v_{\perp}$.
Figure \ref{fig01} shows the radial dependence of $m_{e}|\Psi|$,
which is in units of potential energy (eV) and has been
normalized to zero at the lower boundary of the model.

The \citet{SO79} model is used to compute an iterated electron VDF
$f(v_{\parallel},v_{\perp})$ at a test radius $r_0$ under the
assumption that the VDF at all {\em other} radii is given by the
local Maxwellian $f_{\rm M}$.
For each point in a two-dimensional velocity-space grid at $r_0$,
the conservation of energy and magnetic moment allows us to solve
for $v_{\parallel}$ and $v_{\perp}$ at any other radius $r$.
Some electrons undergo turning points, and in those cases it is
necessary to also evaluate the lowermost and uppermost radii
($r_{\rm L}$ and $r_{\rm U}$, respectively) that are reached by
the electron in question.
When no turning point exists in a given direction, we set either
$r_{\rm L}$ to the lowermost grid zone (1.003 $R_{\odot}$) or
$r_{\rm U}$ to the uppermost grid zone (215 $R_{\odot}$) as needed.

Once an electron's history and bounding radii are known,
the \citet{SO79} collisional optical depth quantity $S$ can be
calculated for all accessible radii
$r_{\rm L} \leq r_{0} \leq r_{\rm U}$, with
\begin{equation}
  S(r)\, =\, \left| \int_{r_0}^{r}
  \frac{dr'}{2 \tau(r') v_{\parallel}(r')}\right| \,\, .
  \label{eq:Sdef}
\end{equation}
By definition, $S(r_{0})=0$, and it increases monotonically in
both directions as one moves away from this evaluation radius.
Distant locations at which $S(r) \gg 1$ are collisionally ``opaque''
and thus unlikely to influence the VDF at $r_0$.
Equation (\ref{eq:Sdef}) utilizes the speed-dependent collisional
timescale,
\begin{equation}
  \frac{1}{\tau} \, = \, \frac{\nu_{ep} \, w_{e}^{3} \, J(w)}{w^3}
  \label{eq:taudef}
\end{equation}
where $w$ is the electron's speed in the solar wind frame,
$\nu_{ep}$ is the classical \citet{B65} electron--proton
collision rate, and $J(w)$ is a dimensionless function that
describes the onset of Coulomb runaway for $w \gtrsim w_{e}$
\citep[see Equation (10) of][]{SO79}.
Figure \ref{fig01} shows the radial dependence of $\nu_{ep}$
to illustrate the rapid loss of collisionality in the low corona.

Note that the explicit radial dependence given for many of the above
quantities conceals the fact that there is an implicit dependence
on the full velocity-space trajectory of the electron in question.
In other words, quantities like $\tau(r)$ and $S(r)$ should be
specified more precisely as, e.g.,
$\tau[v_{\parallel}(r,r_{0}),v_{\perp}(r,r_{0})]$,
and these functional dependences are unique to each
``starting point'' in velocity space.
The optical depth $S(r)$ is a key ingredient in the analogue of the
formal solution to the equation of radiative transfer, which
\citet{SO79} express as
\begin{displaymath}
  f(r_{0}) \, = \, p_{\rm L} \left(
  f_{\rm M}(r_{\rm L}) e^{-S(r_{\rm L})}
  + \int_{r_{\rm L}}^{r_0} dr \,
    \frac{f_{\rm M}(r) e^{-S(r)}}{\tau(r) v_{\parallel}(r)}
  \right)
\end{displaymath}
\begin{equation}
  +\, p_{\rm U} \left(
  f_{\rm M}(r_{\rm U}) e^{-S(r_{\rm U})}
  + \int_{r_0}^{r_{\rm U}} dr \,
    \frac{f_{\rm M}(r) e^{-S(r)}}{\tau(r) v_{\parallel}(r)}
  \right)\,\, .
  \label{eq:ffull}
\end{equation}
\citet{SO79} derived values for the collisional probability
factors $p_{\rm L}$ and $p_{\rm U}$ that are different for the
upward and downward propagating halves of the VDF.
For $v_{\parallel} > 0$, $p_{\rm L}=6/7$ and $p_{\rm U}=1/7$.
For $v_{\parallel} < 0$, $p_{\rm L}=1/7$ and $p_{\rm U}=6/7$.
The use of these values
results in an unphysical discontinuity at $v_{\parallel}=0$, but
it remains a useful first attempt at taking account of the
diffusive nature of collisional transport.

\begin{figure}
\epsscale{1.15}
\plotone{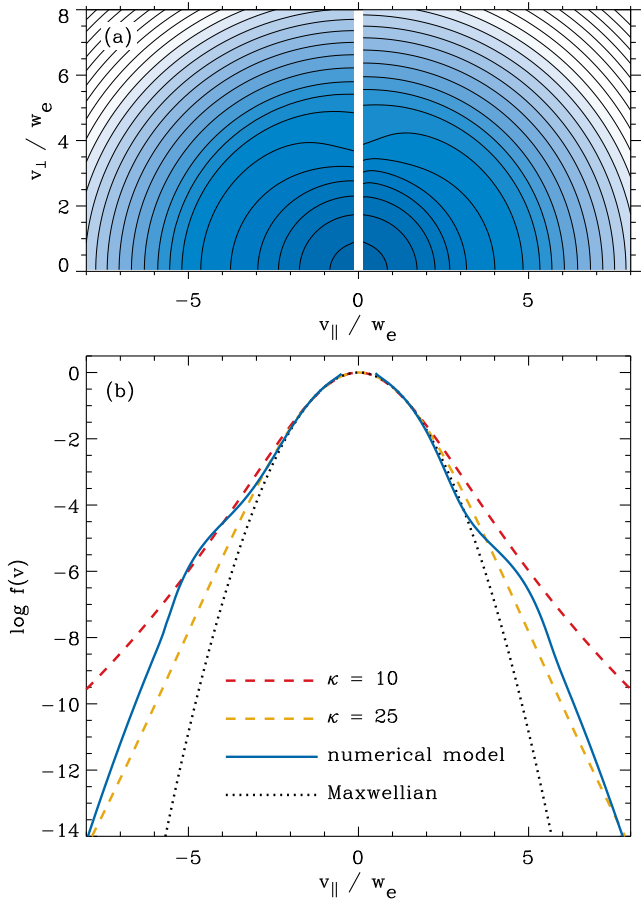}
\caption{(a) Contours of constant $f(v_{\parallel},v_{\perp})$
for the test model at $r_{0}=1.02\, R_{\odot}$, with
velocity coordinates expressed in units of the local thermal
speed $w_e$.
VDF contours are separated by constant factors of 1 in $\log f$.
(b) Slice of $f(v_{\parallel},v_{\perp})$ along the
$v_{\perp}=0$ axis (blue solid curve), compared with the
local Maxwellian $f_{\rm M}$ (black dotted curve) and two
kappa distributions computed with $\kappa=10$ (red dashed curve)
and $\kappa=25$ (gold dashed curve).
\label{fig02}}
\end{figure}

Figure \ref{fig02} shows an example calculation of
$f(v_{\parallel},v_{\perp})$ at $r_{0}=1.02\, R_{\odot}$.
This height, not far above the transition region, is representative
of the location at which the freezing-in of the O$^{+7}$/O$^{+6}$
ratio is expected to occur.
The numerical grid in velocity space was chosen to have 200 points
in $v_{\parallel}$ and 100 points in $v_{\perp}$.
In radial distance, the \citet{CvB07} coronal hole model was
interpolated onto a finer grid of 93,000 points distributed
logarithmically between $r = 1.003 \, R_{\odot}$ and 1~AU.
With such a fine grid, the integrals in Equations (\ref{eq:Sdef})
and (\ref{eq:ffull}) converged well with only first-order
Eulerian quadrature steps.

The discontinuity between the $v_{\parallel}>0$ and $v_{\parallel}<0$
regions of velocity space can be seen most acutely in the mild
suprathermal wings ($v\sim 5w_e$) that arise because of downward
heat transport from the hotter peak temperature at
$r\approx 2\, R_{\odot}$.
The VDF for downward flowing electrons is enhanced relative to
that for upward flowing electrons because of this same heat
transport effect.
The suprathermal enhancement is reminiscent of the hot ``halo''
seen at 1~AU \citep{Fe75}, which is similarly believed to be the
result of nonlocal transport away from the peak temperature region.
At $r_0$, only a small fraction of the total number of electrons
participate in this hot component because the outer corona is
somewhat ``optically thick'' in most parts of velocity space.

For comparison, Figure \ref{fig02}(b) also shows two realizations
of the so-called kappa, or generalized Lorentzian distribution,
\begin{equation}
  f_{\kappa}(v)\,\propto \,\left[ 1+
  \frac{v^2}{(\kappa-3/2) w_{e}^2}\right]^{-1-\kappa}
\end{equation}
\citep[for alternate definitions, see also][]{V68,Cr98,PL10}.
No single value of $\kappa$ fits the numerically computed
VDF, but the two values of $\kappa=10$ and 25 appear
to bracket most of the suprathermal enhancement at $r_0$.

Note that the suprathermal VDF shown above was obtained from the
fastest wind-speed (and lowest density) model of \citet{CvB07}.
To assess the applicability of this result to other solar wind
conditions, we also computed another set of VDFs
using the slowest wind-speed (and highest density) model of
\citet{CvB07}, corresponding to an active-region streamer.
Because this model has a larger transition region height than
the polar coronal hole model \citep[see Figure~17 of][]{CvB07},
we computed the \citet{SO79} model at a correspondingly larger
radius of $r_{0} = 1.03\, R_{\odot}$.
The resulting VDF exhibits a slightly stronger downward-conducting
``shoulder'' at $v_{\parallel}/w_{e} \approx -4$ and a slightly
less intense tail at $|v_{\parallel}/w_{e}| \gtrsim 6$, but most
of the suprathermal electrons still fall between the
$\kappa = 10$ and 25 curves.

\section{Frozen-In Ionization Fractions at 1 AU}
\label{sec:freeze}

\citet{OS83} and \citet{Bu87} first studied the possibility
that suprathermal electrons in the corona may enhance the
frozen-in ion charge state ratios measured at 1~AU.
Later, when it was found that the freezing-in temperatures
are anticorrelated with wind speed \citep[e.g.,][]{vS00},
it was realized that these nonequilibrium ionization processes
could be key diagnostics of the physical processes responsible
for solar wind acceleration.
Thus, in order to better test the MHD turbulence paradigm
used in the \citet{CvB07} models, we want to estimate the
O$^{+7}$/O$^{+6}$ ionization fraction ratios that would be
consistent with the suprathermal VDFs described above.

Figure~15 of \citet{CvB07} showed the wind speed dependence
of the modeled O$^{+7}$/O$^{+6}$ fraction at 1~AU for a series of
18 open flux-tube models of coronal holes, quiescent equatorial
streamers, and active-region streamers.
The original nonequilibrium ionization calculations assumed that
the coronal electron VDFs remained perfectly Maxwellian.
For this paper, a representative freezing-in radius was
determined---in each of these 18 models---by finding the location at
which {\em local} ionization equilibrium would have given the same
O$^{+7}$/O$^{+6}$ ratio as in the full nonequilibrium model.
The temperature $T_f$ at this radius was then assumed to be
that model's core freezing-in temperature, and $T_f$ was
used when looking up the ionization fractions from tables
computed with varying $\kappa$ exponents.
We used the tabulated ionization balance calculations of
\citet{DD13}, which included collisional ionization, autoionization,
radiative recombination, and dielectronic recombination.
This process was also repeated using the earlier kappa-dependent
ionization balance data from \citet{Wn03}, and the results were
the same.

\begin{figure}
\epsscale{1.15}
\plotone{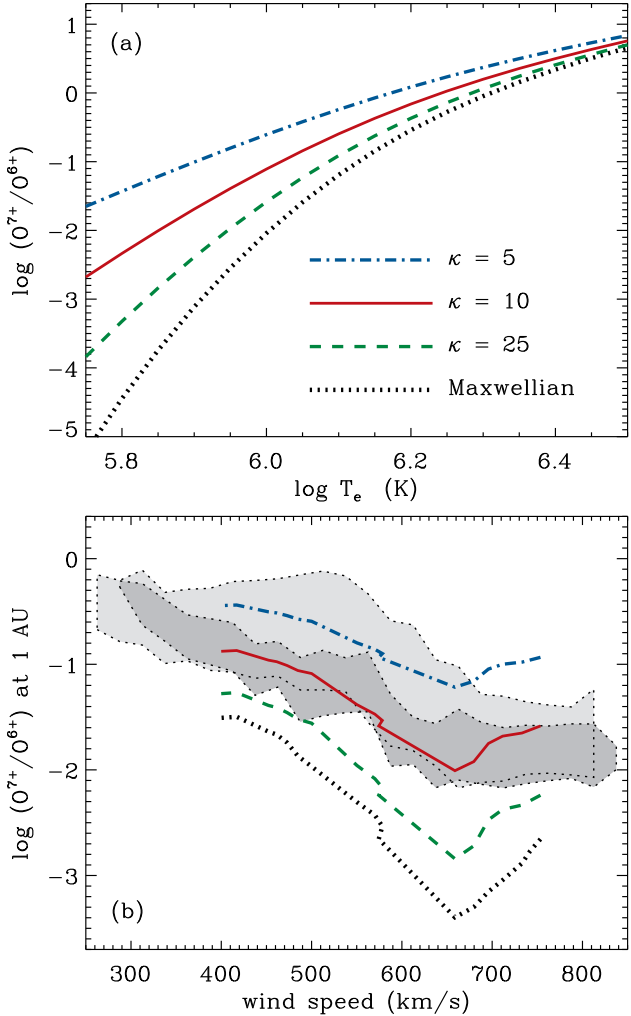}
\caption{Ratio of O$^{7+}$ to O$^{6+}$ number densities plotted
versus (a) equilibrium $T_e$, and (b) solar wind speed at 1~AU
for the standard set of \citet{CvB07} models.
Line styles (consistent in both panels) denote assumed VDF shapes.
Binned {\em Ulysses} data from 1990--1994 solar maximum
({\em light gray region}) and 1994--1995 fast latitude scan
({\em dark gray region}) are shown for comparison.
\label{fig03}}
\end{figure}

Figure \ref{fig03}(a) shows a subset of the equilibrium 
O$^{+7}$/O$^{+6}$ ratios from \citet{DD13}, and
Figure \ref{fig03}(b) shows how these map onto the open
flux-tube models of \citet{CvB07}.
The corresponding ionization fractions for Maxwellian
VDFs were taken from version 7.1 of the CHIANTI database
\citep{Dr97,Ln13}.
The observational data are the same statistical summaries of
{\em Ulysses} SWICS
\citep[Solar Wind Ion Composition Spectrometer;][]{Gl92}
measurements that were presented by \citet{CvB07}.
The model with $\kappa=10$ clearly matches the
SWICS data better than the Maxwellian model.
Figure \ref{fig02}(b) above indicates that this degree of 
suprathermal electron enhancement agrees reasonably well
with (at least the down-conducted half of) what the
\cite{SO79} transport model predicts should exist at the
freezing-in height.

\section{Discussion and Conclusions}
\label{sec:conc}

It has sometimes been asserted \citep[e.g.,][]{An11} that
the striking differences in ion composition between fast and slow
streams is evidence that the two types of solar wind cannot be
driven by the same physical process.
However, the wave/turbulence model of \citet{CvB07} serves as
one counterexample, in which the observed trends in the
O$^{+7}$/O$^{+6}$ and Fe/O (elemental abundance) ratios
versus wind speed are a natural by-product of a single mechanism
operating in differently shaped magnetic flux tubes.
These models utilize identical photospheric lower boundary
conditions, but their variable rates of coronal heating depend on
the magnetic field via the reflection and cascade of Alfv\'{e}n waves.
Conduction from the corona to the transition region
carries this ``information'' back down to the heights at which
different rates of ionization and elemental fractionation occur.
This letter's refined model of suprathermal electron production and
nonequilibrium ionization shows that the wave/turbulence model can
explain not only the observed wind-speed trends, but also the
absolute values of the O$^{+7}$/O$^{+6}$ ratios.

Despite these successes, there is still uncertainty about the ability
of a single type of solar wind heating mechanism to explain the full
range of observed ion composition effects in the heliosphere.
For example, the slow wind from unipolar pseudostreamers
appears to defy the well-known empirical anticorrelation between
wind speed and superradial flux-tube expansion \citep{Wa12}.
Models that employ specific physical processes need to be constructed
for global, three-dimensional descriptions of the heliosphere
\citep[e.g.,][]{vd14,Us14} at times when the coronal and
heliospheric plasma state is well-observed.

Future work must also involve more physical realism for the
models of suprathermal electron transport and nonequilibrium
ionization.
The \citet{SO79} model used above was applied only for a single
iterative step of refinement away from an assumed Maxwellian VDF
in the regions surrounding the test radius $r_0$.
It is suspected that the strength of the suprathermal tails may
be enhanced as a result of iterating multiple times to a
self-consistent set of VDFs over a range of coronal radii.
Improved techniques of describing weakly collisional particle
transport (e.g., solving Fokker-Planck type equations) have been
successful in modeling various suprathermal electron effects in
the corona and solar wind \citep[e.g.,][]{LL00,Vo08,Sm12}.
A full set of nonequilibrium, non-Maxwellian ionization
balance calculations should also be performed for all of the
ions with number densities measured in interplanetary space,
not just O$^{+7}$ and O$^{+6}$ \citep[see][]{Ln12a}.

\acknowledgments

The author gratefully acknowledges Jack Scudder, Ruth Esser,
and John Raymond for many valuable discussions.
This work was supported by NASA grants {NNX\-10\-AC11G}
and {NNX\-14\-AG99G}, and NSF SHINE program grant AGS-1259519.

\end{document}